\documentstyle[12pt,epsfig]{article}
\begin{document}

\vspace*{-0.2in}
\begin{flushright}
CDF/DOC/MUON/PUBLIC/6105 \\
\today\\
\end{flushright}
\vspace{0.8in}

\begin{center}
{\Large {\bf{The "miniskirt" counter array at CDF-II} } }
\end{center}

\vspace{0.6in}
\begin{center}
A.Artikov\footnote[1]{{JINR, Dubna, Russia}},
G.Bellettini\footnote[2]{{INFN, Pisa, Italy}},
J.Budagov$^1$,
F.Cervelli$^2$,
I.Chirikov-Zorin$^1$,
G.Chlachidze$^1$,
D.Chokheli$^1$,
D.Dreossi\footnote[3]{{INFN, Trieste, Italy}},
M.Incagli$^2$,
A.Menzione$^2$,
G.Pauletta\footnote[4]{{University and INFN of Udine, Italy}},
A.Penzo$^3$,
O.Pukhov$^1$,
A.Scribano$^2$,
A.Stefanini$^2$
\end{center}

\renewcommand{\baselinestretch}{1.5}

\vspace*{0.5in}

\begin{center}
\bf{Abstract}
\end{center}

As a part of the CDF Upgrade for Run-II of the Tevatron collider,
the azimuthal coverage of the muon detectors between pseudorapidities of
0.6 and 1.2 was completed by the insertion of stacks of drift chambers and
scintillation counters, which came to be known as the ``miniskirts'' because
they cover lower 90$^0$ in azimuth.

The design and construction of the miniskirts was rather complex and posed
special problems because of its interference with the floor and the supports
of the central detector. The original design parameters of the "miniskirt"
scintillator counters for the CDF Muon System are presented and the
modifications, testing and installation of these counters in the course of
the CDF Upgrade Project are described in detail.

\clearpage

\section{Introduction}

Muon detection is of prime importance for many of the most interesting physics studies
at CDF. In the course of Run-II, the collaboration expects to collect hundreds of
$t\overline{t}$ decays
yielding a muon as well as several million B-hadron events involving
$J/\Psi\rightarrow\mu^+\mu^-$ decays. Muon detection is also of fundamental importance in
the study of {\it W}-boson properties and in the search for {\it Higgs} production associated
with {\it W}- or {\it Z}-bosons~\cite{tdr}. Considerable effort therefore went into extending
the muon detector coverage for Run-II, which started in March 2001.

The CDF-II muon detector system consists of multiple layers of drift chambers and
scintillation counters, which cover the range of pseudorapidities ($|\eta|$) between 0 and 1.5.
Detectors spanning different ranges have different geometries and the muon
scintillation counter system includes subsystems in the regions, which have come to be known
as the ``central'' ($0<|\eta|<0.6$), ``extended'' ($0.6<|\eta|<1.2$) and ``intermediate''
($1.2<|\eta|<1.5$) regions of the detector. During Run-I, the ``extended'' region (referred
to by the acronym CMX) covered only three quarts of the azimuthal acceptance with 8 layers
of drift tubes sandwiched between two layers of scintillation counters (known as the CSX)
\cite{csxnote}. Though much of the missing 90$^0$ of the azimuthal coverage, on both the
east and west sides of the detector, were constructed before Run-I, installation was delayed
until  the ``CDF Upgrade'' and, at that time, it was found necessary to introduce several
modifications to the original design. These subsystems of the CMX came to be known as the
``miniskirts'' because they covered the lower section of the azimuthal range. The counters
(which we refer to collectively as the MSX) of one of these miniskirts are illustrated in
Fig.~1.

In this note we recall the original design parameters and describe in detail the modifications,
testing and installation of the MSX counters in the course of the CDF Upgrade Project.

\section{Design parameters}

Like the CSX, the MSX are trapezoidal counters of NE114 scintillator, produced about 10 years
ago by {\it Nuclear Enterprises}~\cite{ne114}. However, the MSX are 15~mm thick (as opposed
to the CSX, which are 20~mm thick) and the scintillator turned out to be of better quality
than that used for the CSX and for the central counters (known as the CSPs) in that it has
not yet shown any evidence of the premature aging, which has shown up in these latter
counters~\cite{csxnote,aging}. The dimensions of the CSX and MSX are shown in Fig. 2.

The CSX~\cite{csxnote} counters sandwich the drift chambers as shown in Fig.~2a. The
``internal'' (i.e. those closest to the interaction point) and the ``external'' counters are
read out of opposite sides and fed to {\it meantimers} in order to reduce the time variation
associated with hit position~\cite{csxnote}. Due to space restrictions the MSX cover only one
side (that closest to the interaction point) of the miniskirts and the counters are read out
of both ends, through curved Lucite light guides coupled to 2$^{\prime\prime}$
EMI-type~\cite{emi} photomultipliers (PMTs), as illustrated in Fig.~2b. However, additional
space restrictions, not foreseen in the original design did not allow the Lucite light
guides to be installed on one end of four counters (referred to as the MSX$^\prime$) on
each side of each of the two miniskirts (see Fig.~1). Conventional readout was therefore
substituted by wavelength-shifter ($WLS$) fiber readout as illustrated in Fig.~2c. The $WLS$
fibers (1 mm-diameter Y11 (200 ppm) fiber produced by $Kuraray$) were assembled into 15-fiber
ribbons, which were glued (using BC-600 optical cement) to one of the long edges of the
scintillators and read out by means of $Hamamatsu$ H5783 photomultiplier assemblies which
incorporate new ultra compact photomultiplier R5600. Light collection was augmented by
``mirroring'' the end of the fibers furthest from the PMT photocathode. ``Mirroring''
was obtained by evaporating an Aluminium coating on the polished ends of individual fibers.
Reflection coefficients ranging between 70\% and 90\% were obtained in this manner.

This solution was arrived at after preliminary studies of the {\it mean timing} resolution.
Issues related to time resolution are discussed in section 5.

\section{Counter calibrations and performance parameters}

Phototubes were ``calibrated'' by illuminating them with low-level light pulses from a fast
blue LED ({\it NICHIA} NSPB310A), as illustrated schematically in Fig.~3a. The peak
corresponding to the single photoelectron was used to extract the calibration constant {\it k}
(in channels/photoelectron) corresponding to the specific PMT and ADC combination, according
to a procedure similar to that described in \cite{pmtcalib}.  These parameters, as well as
the average number of photoelectrons $\mu$ used for the calibration, the HV at which the
calibration was performed, and the noise rate at that HV, are listed in Table~1 for a pool of
EMI PMTs on hand. Of these, 20 PMTs were selected for low noise, high gain and, for counters
with light-guide readout on both sides, similar characteristics. The remaining 50 PMTs were
new [5].

After calibration and assembly, the number of photoelectrons n$_{pe}$ at the photocathode,
corresponding to the passage of a cosmic muon, which we refer to here as the ``counter
sensitivity'', were measured for all counters. These were obtained from the ADC measurements
of the integrated signals as follows:
\begin{equation}
n_{pe}=\frac{\overline{Q}-Q_{o}}{k/A},
\end{equation}
where $\overline{Q}$ is the mean of the ADC output distribution, $Q_{o}$ is the pedestal,
$k$ is the calibration constant (in channels/photoelectron), for the specific PMT and ADC
combination and $A$ is the amplification factor necessary for isolating the single
photoelectron peak while calibrating.

The characteristics of all MSX counters are listed in Table~2, together with their operating
plateau voltages and, in some cases, dark currents at those voltages. Noise rates
(not all) corresponding to 15 mV discriminator settings are also shown for a set of counters.
These plateau voltages were obtained on the bench, using cosmic muons prior to installation.

A typical light yield distribution corresponding to cosmic muons is shown in Fig.~3b. On the
average, the counter sensitivity is $\sim$80 p.e./muon.

\section{Counter assembly}

All counters were wrapped with diffuse, 0.005$^{\prime\prime}$-thick, aluminized paper
reflector and a combination of 0.015$^{\prime\prime}$-thick black PVC sheet and
0.007$^{\prime\prime}$-thick black electrical tape. Single sheets of the reflector and the
black PVC sheet were successively laid on all flat scintillator faces, which were then taped
down at the edges so the overall thickness of the wrapping material covering the scintillator
was 0.054$^{\prime\prime}$ except for $\sim$1/2$^{\prime\prime}$ along the edges where the
electric tape overlapped the PVC sheet.

All 2$^{\prime\prime}$ PMTs were coupled to their light-guides using standard CERN mounting
assemblies for the EMI 9814B PMT, and they were surrounded by 0.5 mm-thick mu-metal and 1
cm-thick iron shields. The smaller {\it Hamamatsu} tubes were coupled to the polished ends
of the fiber ribbons via Lucite adapters and mounted on Lucite extensions of the counters
as shown in Fig.~2c. They were then surrounded by an aluminium light shield, which was taped
down to the Lucite extension with black electrical tape. All PMTs were air-coupled to their
respective light-guides/adapter.

\section{Timing considerations}

Given the length of the scintillators, which  range between $\sim$160 and $\sim$180~cm, and
the average speed of light in the scintillator ($\sim$15 cm/ns), the uncertainty in the time
at which a muon reaches the counter edge varies between $\sim$10 and $\sim$12 ns. This is
comparable to the difference in the time a muon takes to reach the counters from the
interaction point and the time a background muon takes to reach the same counters from the
beam pipe upstream of the detector, at some point close to the MSX. During Run-I, the background
of this type~\cite{csxnote} would have overwhelmed the CSX signal had it not been for the
introduction of a {\it mean timer} (MT) circuit, described in ref.~2, to sharpen
the timing and eliminate this type of background. In order to reduce this background, additional
shielding  (two concentric rings of 18$^{\prime\prime}$-thick steel, known as the {\it snout}
and 24$^{\prime\prime}$ {\it donut} between the beam pipe and the CSX and MSX counters~\cite{tdr}),
was introduced as a part of the CDF upgrade, however, the acceptance of the plug calorimeters at
low $P_t$ was also increased, with a corresponding increase of material close to the beam pipe,
and additional material in the form of small $P_t$ calorimeters (known as the {\it miniplugs}) has
been added upstream of the plug calorimeters. Therefore, we estimated the worst acceptable timing
resolution from the MSX to be a standard deviation $\sigma$=3~ns, which was sufficient to extract
the signal from the background during Run-I~\cite{csxnote}.

{\it Mean timing} is based on the assumption that the two ends of the detector are symmetrical.
Under these conditions, if the time $t_a$ at which the signal is emitted from the PMT at one end
of the detector is $t + x$, where $x$ is the distance (in ns) from the muon hit to that end of
the detector, then the corresponding time $t_b$ from the other side of the detector will be
$t + l - x$, where $l$ is the length (in ns) of the detector. The mean of these times then differs
from the hit time $t$ which fluctuates because of the combined fluctuations in the number of
photoelectrons ($n_{pe}$) at the PMT photocathodes and in the multiplication factor of the PMTs.
If the counter is not symmetrical, a systematic uncertainty can also be expected to contribute
to the total uncertainty.

Trapezoidal counters are not symmetrical and the MSX$^\prime$ counters, which are read out through a
conventional light guide on the one end and via a $WLS$ fiber ribbon on the other, are much less so.
One might therefore expect a significant systematic contribution from this asymmetry to the
overall $\sigma$. Furthermore, because the $n_{pe}$ for $WLS$ fiber readout is smaller than for
conventional readout, we can also expect an increase in the statistical contribution to the overall
uncertainty. Our first concern, before adopting this solution, was therefore to ascertain that the
{\it mean timing} resolution obtained under these conditions was adequate.

\subsection{Preliminary Tests}

The first tests were performed in a non-destructive manner by leaving the Lucite light guides on
both ends of an MSX counter and optically coupling a $WLS$ fiber ribbon to one of the counter's long
edges by means of optical grease. The light guides were read out by means of EMI 9814B PMTs whereas
the two ends of the $WLS$ ribbon were read out by {\it Hamamatsu} R4125 PMTs (see Fig.~4). By
combining different readouts one could investigate how the corresponding asymmetries in the light
readout contributed to the mean time resolution. For this investigation, the statistical contribution
was reduced by using a relatively high-intensity pulsed (5 ns half-width) U/V laser~\cite{laser} to
illuminate the scintillator at localized positions corresponding to 2 mm-diameter holes in the
0.001$^{\prime\prime}$ black PVC sheet used to wrap the scintillator (see Fig.~4).

The outputs of each of the four PMTs were discriminated and used to stop TDCs, which had been started
by the laser trigger. Sums and differences of pairs of TDCs corresponding to similar readouts are
plotted in Fig.~5a as a function of position along the long dimension of the counter. The effective
velocity of the light in the scintillator and the $WLS$ fiber were calculated from the differences
and found to be 14.6$\pm$0.3 cm/ns and 14.5$\pm$0.7 cm/ns, respectively. The sums of these outputs
correspond to the mean times. Ideally, they should be constant but Fig.~5a shows significant
variations. In the case of light guides readout a variation of $\sim$0.5 ns is observed between the two
ends of the counter. However, the variation is not linear with displacement from one end and it
reaches a maximum of $\sim$1 ns at $\sim$1/3 of the distance from the wider end of the counter. These
variations are attributed to a combination of the trapezoidal shape of the counter and of the
increase of light acceptance as the light source approaches the light guide. The maximum deviation
in the mean times calculated from the two $WLS$ fiber outputs is similar ($\sim$1 ns) and it appears,
as might be expected, to be approximately linear with displacement along the counter. Statistical
uncertainties are considerably larger in this latter case because of the smaller light collection
efficiency.

Systematics related to the displacement of the light source in the direction transverse to the longer
dimension, i.e. across the detector were also investigated and found to be negligible in the case
of conventional light pipe readout. However, variations of up to 1 ns were observed for the
$WLS$ fiber readout.

Variation in the {\it mean times} obtained when the outputs from dissimilar readouts are combined, are
shown in Fig.~5b. Though statistical uncertainties increase due to the effect of the lower
light-collection efficiency of the $WLS$ fiber readout, the systematic variation in the mean times
does not increase significantly.

The conclusion drawn from these measurements was that the asymmetric readout method would not
significantly influence the overall uncertainty in the mean times. Though the effect on the
statistical uncertainty of the lower light collection efficiency from the $WLS$ readout was apparent,
these preliminary tests could not provide a realistic evaluation of the statistical uncertainty
corresponding to minimally-ionizing particles. Measurements of this uncertainty were made using
cosmic muons and obtained $\sigma$=2.9$\pm$0.2 ns for $similar$, light guide, readouts (this result
is consistent with that reported for the CSX in \cite{csxnote}), and $\sigma$=6.3$\pm$0.5 ns for
$dissimilar$ readouts. Given, however, that we expected to increase the light collection efficiency
considerably by ``mirroring'' and by improved optical coupling, we expected that statistical
uncertainties would be reduced to acceptable values when these improvements were introduced.
This expectation was borne out by the tests described below.

\subsection{Timing measurements on the final version of the MSX$^\prime$ \\ counters}

Timing measurements on the MSX$^\prime$ counters, constructed according to the final design
parameters outlined in section 2, were performed with cosmic muons using a scintillator
``telescope'' as illustrated schematically in Fig.~6 (a more detailed description of the data
acquisition (DAQ) electronics is shown in Fig.~7). Two 15x15x2 cm$^3$ scintillators were located
above and below the counter being tested. The coincidence of these scintillators was used to
trigger the DAQ and to start the TDCs which were stopped by the counter outputs. The contribution
$\sigma_{tr}$ of the trigger counters to the total timing uncertainty was measured to be 1.1 ns.
Mean times were calculated by summing the TDC outputs corresponding to the two ends of the detector.
They were also measured directly using a $LeCroy$ 624 meantimer. No significant difference was found
between calculated and measured mean times.

Measurements were performed at 3 different positions along the counters: the center (referred to
as X2), a position (X1), closer to the wider end, at 77 cm from the center, and another position
(X3), closer to the narrower end, at 82 cm from the center. The results of these measurements are
reported in Table 3, where both the full-width at half-maximum (FWHM) and the standard deviation
$\sigma$ of individual times and mean times are tabulated together with the mean times $<T>$ for
all but the shortest (counter 1) of the MSX$^\prime$ counters. The standard deviation $\sigma$ was
corrected for the contribution of the scintillator telescope by subtracting it in quadrature:
$\sigma=\sqrt{\sigma^2_{FWHM}-\sigma^2_{tr}}$.
It was noted that values of the standard deviation were generally higher at X3 (the end furthest
from the PMT) than at other positions. This might be attributed to the presence of Cherenkov light
produced by the passage of the cosmic muons through the light guide, which overlaps this region.

The maximum systematic variations in the mean times $<T>$ are seen to be compatible with those
measured during the preliminary tests. In order to evaluate their effect on the overall uncertainty,
the following algorithm was adopted:
\begin{equation}
FWHM_{tot}=|<T_{X1}>-<T_{X3}>|+\frac{FWHM_{X1}+FWHM_{X3}}{2},
\end{equation}
as illustrated in Fig.~8. $\sigma_{tot}$ was then calculated from $FWHM_{tot}$ by dividing by 2.35.
From these values (tabulated for each counter in Fig.~8), it is apparent that $\sigma_{tot}$ does
not exceed 2.2 ns. This is quite compatible with the requirement that the resolution be sufficient
to discriminate against background in conditions similar to those of Run I~\cite{csxnote}, as
outlined in section 5.

\section{Installation and operation}

Teflon guides and steel runners were then installed on each counters so they could be slid into
position along steel guides fixed to the corresponding drift chamber assemblies. The disposition
of counters, power supply and readout between the collision hall and the CDF electronics room are
shown schematically in Fig. 9.

Power is supplied to the EMI-type PMTs by power supplies, known as ``Gamma Boxes''~\cite{gammabox}
(located in the ``counting room''), via 40-channel distribution units, known as
``Pisa Boxes''~\cite{pisabox} (located close to the detector), through RG58 HV cables.
The $Hamamatsu$ PMTs received their power from supply and distribution units known as CCUs~\cite{ccu},
located close to the detector. All power distribution is controlled and monitored by custom
software~\cite{slow}. This software allows for channel-by-channel control and monitoring, both
manually and automatically, on a regularly scheduled basis.

Analog signals from the EMI-type PMTs are routed to $LeCroy$ 4416 discriminators, located in the
counting room, through 220 ft RG58 cables. All discriminator thresholds are set to their minimum
values (15 mV). Discriminator outputs are split and routed both to meantimer units (located close
to the discriminators) and to TDCs (via 40 ft twist and flat cable). Meantimer outputs are also
routed to TDCs via the same twist and flat cables.

Presently (fall 2002), only three sections of standard (i.e. light-guide readout on both sides)
are presently installed. The remaining (North-West) section and the four wedges with the
MSX$^\prime$ counters will probably be installed during a forthcoming shutdown.

An example of regularly monitored ``occupancy plots'', recorded by the YMON program, is shown in
Fig.~10 for both beam and cosmic data.


\begin{thebibliography}{99}
\bibitem{tdr}  ''The CDF-II Detector Technical Design Report''.
The CDF II collaboration, Fermilab-Pub-96/390-E (1996).

\bibitem{csxnote}  P.~Giromini et al. ''The Central Muon
Extension Scintillators (CSX)''. CDF Note 3898 (1996)

\bibitem{ne114} Supplied in 1992/93 by {\it Nuclear Enterprises Co}.
Equivalent scintillator is now produced by $Bicron$ (BC416).

\bibitem{aging}  S.~Cabrera et al., ''Making the Most of Aging
Scintillator'', Nucl.Instrum.Meth. A453 (2000) 245-248

\bibitem{emi} Original PMTs were EMI 9814B PMTs. These PMTs
are now produced by {\it Electron Tubes Inc.} (UK). Fifty new PMTs were used for
the Miniskirt couters. The remaining 20 were older EMI 9814B PMTs.

\bibitem{pmtcalib} E.N.~Bellamy et al. ''Absolute calibration
and monitoring of a spectrometric channel using a photo-multiplier''.
Nucl.Instrum.Meth. A339 (1994) 468-476.

\bibitem{laser} LN300 pulsed Nitrogen laser produced by {\it Laser Photonics Inc.}
It features a spectral output of 337.1 nm, pulse width of 300 ps, energy stability
of 5\%.

\bibitem{gammabox} Supplied by {\it Gamma High Voltage Research Inc.} (US).

\bibitem{pisabox} Pisaboxes originally built at University of
Pisa. Electrical engineers who originally designed Pisaboxes left and
founded CAEN (Costruzioni Apparecchiature Elettroniche Nucleari). Pisaboxes
are a very simple idea: use small motors connected to potentiometers to set
HV, and control and read back values with a microprocessor (8080 vintage).
Communications to the Pisabox was a serial connection daisy-chained via
RG-174 from Pisa-box controller CAMAC module (''Pisabox protocol''). This
was improved first to CAENet, and then H.S. CAENet, which is used on CAEN
527 HV supplies.

\bibitem{ccu} C.~Bromberg for the CDF Collaboration, ''Gain and
Threshold Control of Scintillation Counters in the CDF Muon Upgrade for
Run-II'', Int.Journal of Mod.Phys.A Vol.16, Suppl. 1C (2001) 1143-1146.

\bibitem{slow} O.~Pukhov et al. ''Automatization of the Muon
Scintillator System Power Supply at CDF-II'', CDF Note 5949 (2002).
\end{thebibliography}
\end{document}